\documentclass[smallextended]{svjour3}       % onecolumn (second format)
\smartqed  % flush right qed marks, e.g. at end of proof
\usepackage{graphicx}
\usepackage{amsmath}
\usepackage{amssymb}
\newcommand{\Op}{\mathcal{O}}
\newcommand{\br}[1]{\langle #1 \rangle}
\newcommand{\bra}[1]{\langle #1 |}
\newcommand{\ket}[1]{| #1 \rangle}
\newcommand{\e}{\mathrm{e}}

\begin{document}

\title{Recent progress on cluster and meron algorithms for strongly correlated systems}
\author{Debasish Banerjee}
\institute{Saha Institute of Nuclear Physics, HBNI, 1/AF Bidhannagar, Kolkata 700064, India.\\
\email{debasish.banerjee@saha.ac.in}}

\date{Received: date / Accepted: date}

\maketitle

\begin{abstract}
Ab-initio studies of strongly interacting bosonic and fermionic systems is greatly facilitated by efficient
Monte Carlo algorithms. This article emphasizes this requirement, and outlines the ideas behind the
construction of the cluster algorithms and illustrates them via specific examples. Numerical studies of
fermionic systems at finite densities and at real-times are sometimes hindered by the infamous sign problem,
which is also discussed. The construction of meron cluster algorithms, which can solve certain sign problems
are discussed. Cluster algorithms which can simulate certain pure Abelian gauge theories (realized as quantum
link models) are also discussed.

\keywords{Cluster Algorithms \and Fermionic and Spin models \and Lattice Gauge Theories}
%\PACS{PACS code1 \and PACS code2 \and more}
\end{abstract}

\section{Introduction: why cluster algorithms?} \label{intro}
The ab-initio studies of strongly correlated systems occurring in Nature, whether in particle physics or in
condensed matter physics, is an extremely challenging topic. Analytical solutions are difficult to find in
most cases, and perturbation theory fails for strong couplings. Certain weak coupling methods (such as the
epsilon-expansion \cite{Card2008,Peli2002}) have been successful in addressing the existence of fixed points
in renormalization group flow, as well as compute critical exponents at phase transitions by systematically
improving over the mean field estimates. Large-N methods have provided another analytic handle on some
interesting quantum field theories (QFTs) \cite{Mosh2002}. In the case of conformal field theories (CFTs)
there has been exciting developments through the use of AdS/CFT \cite{Mald1998}, and the more recent
conformal bootstrap \cite{Rata2008} and the large charge expansion \cite{Hell2015}. Tensor Networks
\cite{Scho2011} have significantly contributed to computing paradigms in lower dimensional systems, by
opening up the possibility to simulate a large range of strongly interacting systems, even in real-time.

However, in the overwhelmingly large majority of cases, the Markov Chain Monte Carlo (MCMC) methods, starting
from the initial proposal of Metropolis et al. \cite{Metr1953} have provided an unbiased ab-initio method to
numerically sample multidimensional integrals and evaluate the expectation values of physical operators. Let
us denote the degrees of freedom in a (classical or quantum) system by $\{ q_i \}, i = 1, \cdots, V$, and the
partition function for the system as
\begin{equation}
  {\cal Z} = \int {\cal D} q~ \e^{-S(\{q_i\})} = \mathrm{Tr~~exp} (-\beta H) ,
\end{equation}
where $S(\{q_i\})$ is an action functional corresponding to the Hamiltonian $H$ at an inverse temperature $\beta$.
The quantity $\mathrm{e}^{-S(\{q_i\})} = W(\{q_i\})$ is the Boltzmann weight and is typically positive for the
chosen computational basis of $\{ q_i \}$. If this is not the case, we encounter the {\bf sign problem} and
importance sampling fails. We will assume the positivity of the Boltzmann weight for the moment, and will come
back to the exceptions later. The expectation value of a physical operator $\Op$ (for example an order parameter
or a correlation function) is
\begin{equation}
  \br{\Op} = \frac{1}{{\cal Z}} \int {\cal D} q ~ \Op~ \e^{-S(\{q_i\})}
\end{equation}

Starting from an initial probability distribution, a Markov Chain is a series of probability distributions
$\Pi_k (\{q_i\})$, which steadily converge to the fixed point distribution $\Pi^{\star} (\{q_i\})$, which is the
equilibrium distribution $W(\{q_i\})$. We will assume that the reader is familiar with the basics of Markov Chains,
and local Monte Carlo algorithms and point to excellent textbooks which cover this topic in depth \cite{MC1,MC2}.
While it is not always difficult to construct a Markov Chain with the necessary properties, the difficulty lies
ensuring that the rate of convergence is insensitive to V. Further, even if a Markov chain were to converge to
the equilibrium distribution, one needs to sample this distribution to obtain uncorrelated samples on which
expectation values and correlation functions of relevant operators can be measured. This is exactly where improved
algorithms, such as the cluster algorithm, assert their importance. To make this notion quantitative, we first
introduce the concept of {\bf autocorrelation time}.

Operationally, after equilibrium is reached, (local) Monte Carlo algorithm generates a set of configurations of
$(\{q_i\})$ according to the Boltzmann weight, on which physical operators are measured. Let us denote the
distribution at the $l$-th Monte-Carlo time as $(\{q_i\})_l$. Typically the distributions $(\{q_i\})_l$ and
$(\{q_i\})_{l+1}$ are highly correlated since only a small change is involved in one step to the next. Therefore,
measurement of the physical operator on these two subsequent configurations, $\Op_l (\{q_i\})$ and
$\Op_{l+1} (\{q_i\})$ are also not independent.

Numerical calculations necessarily deal with finite data, let us denote this by $N_{\rm conf}$. The sample mean
of the dataset is $\overline{\Op} = \frac{1}{N_{\rm conf}} \sum_l \Op_l$. According to the central limit theorem,
the sample mean $\overline{\Op}$ has a Gaussian distribution about the exact expectation value $\br{\Op}$, such
that $\br{\Op} = \overline{\Op} \pm \sigma_{\overline{\Op}}$. With uncorrelated measurements, the best unbiased
estimate of the error from the finite data-set is
\begin{equation} \label{eq:err}
  \sigma_{\overline{\Op}} = \left[ \frac{1}{N_{\rm conf} (N_{\rm conf}-1)} \sum_k (\Op_k - \overline{\Op})^2 \right]^{\frac{1}{2}}.
\end{equation}
Note that we have used $\overline{\Op}$ as the best estimate of the true mean, $\br{\Op}$. For large $N_{\rm conf}$,
\begin{equation}
   \sigma^2_{\overline{\Op}} = \left[ \frac{1}{N_{\rm conf}} \sum_m (\Op_m - \overline{\Op}) \frac{1}{N_{\rm conf}} \sum_n (\Op_n - \overline{\Op}) \right] = \frac{1}{N^2_{\rm conf}} \sum_{m, n}^{N_{\rm conf}} \Gamma_{\rm \Op} ( m - n ).
\end{equation}
The autocorrelation function, $\Gamma_{\rm \Op} (m)$, is a property of the Monte Carlo algorithm and is defined through 
the unequal time-correlator
\begin{equation}
  \Gamma_{\rm \Op} (m) = \br{\Op_{m^\prime} \Op_{m + m^\prime}} - \br{\Op}^2 = \Gamma_{\rm \Op} (-m) \overset{m \gg 1}{\sim} C \exp (-m/\tau_{\rm exp})
\end{equation}
and decays exponentially at asymptotic times, with the decay time $\tau_{\rm exp}$ corresponding to the slowest
mode in the Monte Carlo dynamics. The {\bf integrated autocorrelation time}, $\tau_{\rm int}$, defined as
\begin{equation}
  2 \tau_{\rm int} \approx \sum_{m=-M}^{M} \Gamma_{\rm \Op}(m) = \Gamma_{\rm \Op}(0) + 2 \sum_{m=1}^M \Gamma_{\rm \Op} (m)
\end{equation}
and the cut $M$ is due to finiteness of the dataset, and we have ignored corrections of order $\tau_{\rm exp}/N_{\rm conf}$,
for large $N_{\rm conf}$. Note that $\Gamma_{\rm \Op} (0)$ is the sample variance. The expression for the error on
the mean is then
\begin{equation}
  \sigma^2_{\rm \Op} = \frac{2 \tau_{\rm int} \Gamma_{\rm \Op} (0)}{N_{\rm conf}} = \frac{\Gamma_{\rm \Op} (0)}{N_{\rm eff}}
\end{equation}
implying that there are only $N_{\rm eff}$ independent configurations in the Markov chain. Unlike the equilibrium
values, $\tau_{\rm int}$ depends on the algorithm. Near a critical point, for example, autocorrelation times diverge
with the physical correlation length $\xi$ as $\tau \sim c \xi^z$, and $z \sim 2$ for local algorithms ($z$ is called
the dynamical critical exponent). Such studies are therefore numerically challenging especially near the critical
point, and the phenomena is called {\it critical slowing down}. Cluster algorithms can guarantee extremely small
$\tau_{\rm int}$, and sometimes even $z \sim 0$, thereby practically eliminating this problem.

\section{Quantum spin models} \label{sec:1}
Using the formulation of Fortuin and Kasteyln \cite{Kast1969,Fort1972}, the first cluster algorithm for simulating
classical spin (the Ising and the Potts) models were constructed by Swendsen and Wang \cite{Swen1987}, and then
extended by Niedermayer \cite{Nied1988} and Wolff \cite{Wolf1989} for continuous spins systems. We refer to the
review \cite{Nied1997} for more details. Cluster algorithms for continuous systems, such as hard spheres, have also
been developed. We do not discuss them here, but refer to \cite{Krau2003} for a more detailed exposition. We instead
focus on cluster algorithms for quantum systems, starting with quantum spins.

Cluster algorithms for quantum spin systems, in the {\bf world line formulation} of quantum spins, were first
developed in \cite{Ever1993,Ever21993}, and extended to the continuous time version in \cite{Bear1996}. We note
that the Stochastic Series Expansion (SSE) of the quantum spin Hamiltonian as developed in \cite{Sand1991}
leads to a loop Monte Carlo updating method \cite{Sand2002}, similar to the cluster algorithm. There has been
significant developments in generalizing this algorithm for a variety of models, larger quantum spins, and we
refer the reader to \cite{Ever2002} for details and a more complete set of references. We note that the {\bf worm
algorithm} is another powerful idea which has led to the development of efficient algorithm for a large class of
models, and in many cases is as powerful as the cluster algorithms. We refer to the interested reader to 
\cite{Prok2001,Boni2006} for details.

\begin{figure*}
\includegraphics[width=0.4\textwidth]{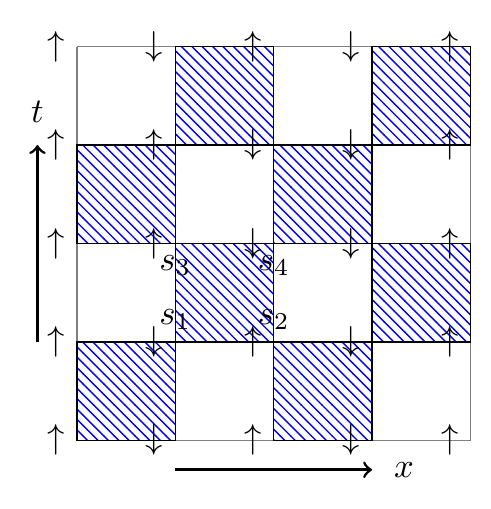}
\vspace{0.8cm}
\includegraphics[width=0.7\textwidth]{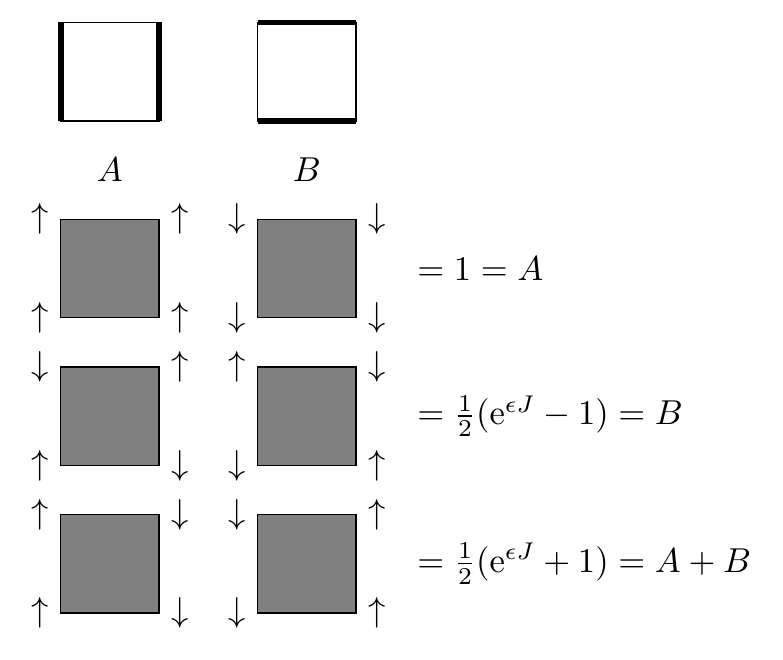}
\caption{({\bf left}) An example of a world-line configuration of the anti-ferromagnetic Heisenberg model. There
is one transition and one anti-transition which spatially displaces a spin-up (down) and brings it back. ({\bf right})
Possible bond-types for this model: (a) the A-type breakup connects identical spins forward in time, (b) the B-type
breakup connects opposite (spin-up with spin-down) spins sideways. We can figure out the probability of the weights
by computing which breakup can be applied for which interaction plaquette. For those where the spin-states do not,
and cannot change (top), only A can be applied, whereas where the spin state flips (middle) only B can be applied.
The last case is where the spin state {\it can} change but did not. Here, the A-breakup can be applied with the
probability $W_A = A/(A+B) = 2/({\rm exp}(\epsilon J) + 1)$.}
\label{fig1}
\end{figure*}

Let us illustrate the construction of a cluster algorithm for the Heisenberg model, which not only has a very
rich physical pedagogy behind it, but is also useful to understand the magnetic properties in certain electronic
systems \cite{Heis1928}. This method is independent of spatial dimensions, but we will consider $(2+1)-$d,
anti-ferromagnetic version $J > 0$ for definiteness. Starting from the Hamiltonian on a square lattice of linear
extent $L$, the partition function ${\cal Z}$ at an inverse temperature $\beta$ is:
\begin{equation}
 {\cal Z} = \mathrm{Tr~~exp} (-\beta H);~~~ H = J \sum_{x,\hat{i}=1,2} \vec{S}_x \cdot \vec{S}_{x+\hat{i}}.
\end{equation}
Using the Suzuki-Trotter formula \cite{Suzu1976}, we separate the Hamiltonian into 4 parts ($2d$ parts in
$d-$spatial dimensions), $H = H_1 + H_2 + H_3 + H_4$, such that the spins contained in each part mutually commute.
\begin{equation}
\begin{aligned}
 H_1 &= J \sum_{x=(2m,n)} \vec{S}_x \cdot \vec{S}_{x+\hat{1}};~~ H_2 = J \sum_{x=(2m+1,n)} \vec{S}_x \cdot \vec{S}_{x+\hat{1}}; \\
 H_3 &= J \sum_{x=(m,2n)} \vec{S}_x \cdot \vec{S}_{x+\hat{2}};~~ H_4 = J \sum_{x=(m,2n+1)} \vec{S}_x \cdot \vec{S}_{x+\hat{2}}.
\end{aligned}
\end{equation}
We construct ${\cal Z}$ by inserting intermediate time-slices, such that $\beta = 4 N \epsilon$, and $\epsilon$
is the temporal lattice spacing. Using ${\cal I} = \ket{n_k}\bra{n_k}$, we expand ${\cal Z}$ as a product over
the matrix elements of the transfer matrix, expressed in the $S^z=\pm \frac{1}{2}$ basis, which is chosen as the
computational basis:
\begin{equation}
\begin{aligned}
{\cal Z} &=  \mathrm{Tr~~exp} (-\beta H) = \lim_{\substack{N \to \infty \\ \epsilon \to 0}} \sum_{n} \bra{n} \left[ {\rm e}^{-\epsilon (H_1 + H_2 + H_3 + H_4)} \right]^N \ket{n} \\
         &= \lim_{\substack{N \to \infty \\ \epsilon \to 0}} \sum_{\substack{\{n \} \\ n_{4N} = n_0}} \prod_{k=0}^{N-1} \bra{n_k} {\rm e}^{-\epsilon H_1} \ket{n_{k+1}} \bra{n_{k+1}} {\rm e}^{-\epsilon H_2} \ket{n_{k+2}} \bra{n_{k+2}} {\rm e}^{-\epsilon H_3} \ket{n_{k+3}} \bra{n_{k+3}} {\rm e}^{-\epsilon H_4} \ket{n_{k+4}}
\end{aligned}
\end{equation}
We have used the Trotter formula in the second line, and the matrix elements connect only the specific bonds. Finally,
we can express the matrix elements via the local action
\begin{equation}
{\cal Z} = \prod_{x,t} \sum_{s(x,t) = \pm 1} {\rm e}^{-S [s_1, s_2, s_3, s_4]}
\end{equation}
where the term $S [s_1, s_2, s_3, s_4]$ connects two neighboring spins, $s_1, s_2$ at a time-slice, $t$, with their
forward-in-time partners $s_3,s_4$. Note that only a set of bonds are {\it active} at a given time-slice, and all
others are passive, and this 4-spin interaction traces an {\bf active plaquette}. A schematic figure for the Trotter
decomposition in $(1+1)-$d is shown in Fig \ref{fig1} (left), where the shaded plaquettes are the active ones, and
indicate the spin pairs which are interacting at that given time. A representative $s_1, s_2, s_3, s_4$ are also
shown in the same figure. For the anti-ferromagnetic Heisenberg model, the explicit values of the transfer matrix
(in the basis where the states of the two spins at sites $x$ and $y$ are $\ket{\uparrow \uparrow}, \ket{\uparrow \downarrow},
\ket{\downarrow \uparrow}, \ket{\downarrow \downarrow}$) are:
\begin{equation}
{\rm e}^{-S[s_1,s_2,s_3,s_4]} = \bra{s_1 s_2} {\rm e}^{-\epsilon J \vec{S}_x \vec{S}_y} \ket{s_3 s_4}  = {\rm e}^{-\frac{\epsilon J}{4}}
  \begin{bmatrix}
 1 & 0 & 0 & 0 \\
 0 & \frac{1}{2}(1 + {\rm e}^{\epsilon J}) & \frac{1}{2}(1 - {\rm e}^{\epsilon J}) & 0 \\
 0 & \frac{1}{2}(1 - {\rm e}^{\epsilon J}) & \frac{1}{2}(1 + {\rm e}^{\epsilon J}) & 0 \\
 0 & 0 & 0 & 1 \\
  \end{bmatrix}
\end{equation}
Note that there are only two off-diagonal elements, both of which are negative. For a bipartite lattice, the negative
sign can be eliminated by doing a unitary transformation on every alternative spin, such that the off-diagonal
elements become $\frac{1}{2}({\rm e}^{\epsilon J}-1)$. For a triangular lattice, for example (or for any non-bipartite
lattice, in general), this does not work, and we encounter the first example of a {\bf sign problem}. In the chosen
basis the probability weights are not positive definite, and Monte Carlo simulation cannot be performed.

Next, to construct a cluster algorithm we have to expand the configuration space by including bond variables, together
with the quantum spins. This implies choosing the bond variables $[b]$ such that the following equation is satisfied
for the different possibilities of $s_1, s_2, s_3, s_4$,
\begin{equation}
  {\rm e}^{-S[s_1,s_2,s_3,s_4]} = \sum_{[b = A,B]} {\rm e}^{-S[s_1,s_2,s_3,s_4; b]}.
\end{equation}
In this particular example, as shown in Fig \ref{fig1}, the linear equations can be satisfied by the use of two bond
breakups, $A$ and $B$. The clusters are constructed as follows: (a) start from an initial site,  (b) follow the chosen
breakups as explained in the caption of Fig \ref{fig1} (right) until the initial point is reached. In this case, the
resulting cluster is the same as a loop, and hence the relation to the loop algorithm. Flipping the cluster implies
the operation $\ket{\uparrow_k} \leftrightarrow \ket{\downarrow_k}$, which preserves the Boltzmann weight of the
configuration, and yet changes the configuration globally, thereby decorrelating them very fast. This is the reason
for the efficiency of the cluster algorithm --- the clusters correspond to correlated degrees of freedom, which get
efficiently updated by flips. In fact, representing the configurations in terms of clusters allows one to 
construct improved estimators for quantities such as the magnetization, susceptibility and the correlation function.

Let us illustrate the construction of improved estimators through examples. The total (uniform) magnetization of the
spins is given as ${\cal M} [\{S^z\}] = \sum_i (S^z)_i = \sum_C {\cal M}_C $, where $i$ runs over all the space-time
lattice sites, and $C$ runs over all the clusters into which a single configuration of spins has been decomposed. The 
equality follows since each spin must uniquely belong to a cluster. The cluster magnetization is thus given as
${\cal M}_C = \sum_{i \in C} (S^z)_i$. Since each cluster flip changes the sign of ${\cal M}_C$, we conclude that 
$\br{{\cal M}} = 0$. Note that this is consistent with the physical result that the magnetization always vanishes in 
a finite volume. Such results are extremely difficult to observe, especially in the broken phase, with local update 
algorithms. Another example is the (uniform) susceptibility:
\begin{equation}
	\chi = \frac{\beta}{V} \br{{\cal M}^2} = \frac{\beta}{V} \br{ ( \sum_C {\cal M}_C )^2 } 
	     = \frac{\beta}{V} \br{\sum_{C_i, C_j} {\cal M}_{C_i} {\cal M}_{C_j}}
	     = \frac{\beta}{V} \br{\sum_{C} {\cal M}^2_C}
\end{equation}
Note that different clusters $C_i$ and $C_j$ are uncorrelated and hence the cross-terms vanish under the operation
of all cluster flips, leaving behind only diagonal terms. Further in a cluster $C$, since all the spins are pointing
in the same direction, the magnetization is given by the total cluster size, ${\cal M}_C = \pm|C|$. This allows us to
rewrite $\chi$ as follows, $\chi = \frac{\beta}{V} \br{\sum_C |C|^2}$. Thus it is evident that clusters are relevant
physical objects, since their size is related to a physical quantity.

Two variants of this cluster algorithm are well-known: {\it multi-cluster} algorithm, which proceeds to construct
all the possible clusters on a given configuration of spins and bonds, and then flips each cluster with a $50 \%$
probability, and the {\it single-cluster}, or the Wolff algorithm which constructs a single cluster at always flips it.
It is also possible to construct this cluster algorithm in continuous time.

It is to be noted that both types of cluster algorithms suffer from {\it Trotter errors}, caused due to the finite 
$\epsilon$ essential in the numerical simulations. In particular, the errors on the
physical quantities go as $\epsilon^2$, and thus one typically simulates for several different $\epsilon$ values to
take the time-continuum limit. It is however possible to construct the cluster algorithms directly in the time-continuum
limit \cite{Bear1996}, which eliminate the Trotter errors completely. 

It must be emphasized that for the cluster algorithm to be successful, certain special configurations called {\bf
reference configurations} need to exist. In the case of the Heisenberg anti-ferromagnet the reference configuration(s)
are the ones where the spins are staggered on the bi-partite lattice. Any configuration can then be decomposed into a
certain set of clusters, and the clusters are accordingly flipped to bring the configuration into (one of) the
reference configurations. This trivially proves ergodicity. In many cases (examples are frustrated magnets, spin
glasses, gauge theories), while it is conceivable to build clusters, absence of reference configurations causes the
clusters to grow too big, almost filling up the whole volume. In such a case, the cluster algorithm does not perform
any better than a local update algorithm.

\section{Fermions} \label{sec:2}
Attempting to simulate fermions brings us head on with the {\bf fermion sign problem}. This is most directly seen
in the world-line representation of the fermions in the occupation number basis. In dimensions $d \geq 2$, it is
easy to construct world-lines of identical fermions which twist around each other as they evolve in Euclidean time.
Configurations which have fermions exchanging positions an odd number of times have an overall negative sign compared
with those where fermions exchange positions an even number of times (everything else being identical), due to the
Pauli exclusion principle. A particularly clear statement of this fermion sign problem and what it entails to solve
the problem is given in \cite{Troy2005}.

As in the case of quantum spins, to simulate fermionic systems we construct the partition function as usual. In the
Lagrangian formulation, one uses Grassmann variables, which is typically integrated out to leave behind a determinant
involving auxiliary fields (if the theory has four-Fermi interactions, or Yukawa couplings), or even gauge fields
(as in quantum electrodynamics, or quantum chromodynamics). Updating procedures either involve updating the determinant
directly \cite{Blan1981,Assa2008}, or recast the determinant in terms of bosonic fields, which are then updated
\cite{Duan1987}. To construct cluster algorithms for fermions, we first obtain a bosonic representation using the
Jordan-Wigner transformation. If the resulting bosonic system can be efficiently updated using a cluster algorithm,
then one can identify types of interactions for which the fermion sign problem can be solved \cite{Chan2002}. When such
an approach succeeds, it is known as the {\bf meron algorithm} first introduced in \cite{Chan1999}, and will be
discussed below. Other novel approaches, such as the {\bf fermion bag} \cite{Chan2010,Huff2017} or the {\bf diagrammatic Monte
Carlo} \cite{Boni2006,Houc2010} are getting increasingly popular in simulating fermionic systems. The former method
expands the fermionic action and groups regions where the fermions can propagate freely against regions where they
are bound into monomers and dimers. The Monte Carlo procedure then updates these regions, which are denoted as fermion
bags. The latter approach directly samples the Feynman diagrams associated with the interactions, and in certain
regimes can be related to the bag approach.

Before discussing the construction of the meron algorithm, we elaborate the nature of the fermionic sign problem,
which will also serve to understand the solution. First, we set up the fermionic path integral in the occupation
number basis $\{n_k\}$:
\begin{equation}
{\cal Z}_{\rm f} = \mathrm{Tr~~exp} (-\beta H_{\rm f}) = \sum_{ \{ n_k \} } p(\{n_k \}); \br{A} = \frac{1}{{\cal Z}_{\rm f}} \sum_{\{ n_k \}} A(\{ n_k \}) p(\{n_k \}),
\end{equation}
where $p(\{n_k \})$ is the probability for a given configuration, and $A$ is an operator in the occupation number
basis. To qualify the severity of this problem, one can consider the sign of the configuration as a part of the
observable $A$ through the decomposition $p(\{n_k \}) = {\rm sign}(\{n_k\}) |p(\{n_k \})|$, the sign being $\pm 1$
depending on whether the fermions in the configuration swap their positions even or odd number of times. It follows
that
\begin{equation}
\begin{aligned}
\br{A}_{\rm f} &= \frac{\sum_{\{ n_k \}} A(\{ n_k \}) p(\{n_k \})}{\sum_{\{ n_k \}} p(\{n_k \})} \\
               &= \frac{\sum_{\{ n_k \}} A(\{ n_k \}) {\rm sign}(\{n_k\}) | p(\{n_k \})|}{\sum_{\{ n_k \}} |p(\{n_k \})|} \times \frac{\sum_{\{ n_k \}} |p(\{n_k \})|}{\sum_{\{ n_k \}} {\rm sign}(\{n_k\}) |p(\{n_k \})|} \\
               &= \frac{ \br{A  \cdot {\rm sign}}_{\rm b} }{ \br{{\rm sign}}_{\rm b}}
\end{aligned}
\end{equation}
The subscript ${\bf b}$ indicates that the averaging is done over a bosonic system whose weights, $| p(\{n_k \}|$,
are positive definite by construction. The quantity $\br{{\rm sign}}_{\rm b}$ measures the severity of the sign problem.
It can shown that $\br{{\rm sign}}_{\rm b} = {\rm exp}(-\beta V \Delta f)$, where $\Delta f = f_{\rm f} - f_{\rm b}$
is the difference in free energy density between the fermion and bosonic ensembles. At low temperatures and large
volumes, this quantity is exponentially small:
\begin{equation}
\frac{(\sigma_{\rm sign})_{\rm b}}{\br{\rm sign}_b} = \frac{\left[ \br{{\rm sign}^2}_{\rm b} - \br{{\rm sign}}^2_{\rm b} \right]^\frac{1}{2}}{\sqrt{N}\br{\rm sign}_b} \approx \frac{{\rm exp}(\beta V \Delta f)}{\sqrt{N}},
\end{equation}
where $N$ is the number of uncorrelated data, and we have used $\br{{\rm sign}^2}_{\rm b}=1, \br{{\rm sign}}_{\rm b} \approx 0$.
Clearly, $N \sim {\rm exp}(2 \beta V \Delta f)$ to determine the sign. Thus even if $\br{A}_{\rm f} \sim 1$, it is
measured as the ratio of two exponentially small signals, and requires exponential effort to extract from statistical
noise.

One could aim to cancel the negative signs by pairing the configurations which carry an $-1$ sign with those that carry
a $1$ sign, such that only a few unmatched configurations remain. In fact, this is the idea behind the meron algorithm.
It is however important to realize that this solves only one-half of the sign problem. With the matching, we
effectively have ${\rm sign}^2 = {\rm sign}$, since ${\rm sign}=0,1$ and hence
\begin{equation}
\frac{(\sigma_{\rm sign})_{\rm b}}{\br{\rm sign}_b} = \frac{\left[ \br{{\rm sign}}_{\rm b} - \br{{\rm sign}}^2_{\rm b} \right]^\frac{1}{2}}{\sqrt{N}\br{\rm sign}_{\rm b}} \approx \frac{{\rm exp}(\beta V \Delta f/2)}{\sqrt{N}},
\end{equation}
Note that we have achieved an exponential gain in statistics, but one still needs $N \sim {\rm exp}(\beta V \Delta f)$
for any meaningful result. The {\bf meron algorithm} incorporates the two steps together: it identifies the
configurations which can be analytically cancelled, {\it and} never generates them. Thus, one always generates
configurations which contribute non-trivially to ${\cal Z}_{\rm f}$. The algorithm, however, only works for a
restricted class of interactions.

Let us illustrate this for the ${\rm t}-{\rm V}$ model in $(2+1)-$d, extensively used in the study of certain electronic
systems (though the algorithm can be constructed in any dimensions). The Hamiltonian is
\begin{equation}
{\rm H}_{\rm f} = -\frac{{\rm t}}{2} \sum_{x,\hat{i}=1,2} ( c^\dagger_x c_{x+\hat{i}} + c^\dagger_{x+\hat{i}} c_x) + {\rm V} \sum_{x,\hat{i}=1,2} (n_x - \frac{1}{2}) (n_{x+\hat{i}} - \frac{1}{2}),
\end{equation}
with $V \geq t > 0$. The fermion creation (annhiliation) operators at site x, $c_x (c^\dagger_x)$ satisfy the following
anti-commutation relations: $\{ c_x, c_y\} = \{ c^\dagger_x, c^\dagger_y\} = 0; ~~\{ c^\dagger_x, c_y\} = \delta_{xy}$.
The corresponding bosonic system is obtained by using the Jordan-Wigner transformation. An ordering of the lattice
points is defined: $ l = x + y \cdot L$ (in $d=2$), and the fermionic operators are expressed as follows:
\begin{equation}
 c^\dagger_x = \sigma^3_1 \cdot \sigma^3_2 \cdots \sigma^3_{l-1} \sigma^+_l;~~c_x = \sigma^3_1 \cdot \sigma^3_2 \cdots \sigma^3_{l-1} \sigma^-_l;~~n_x = c^\dagger_x c_x = \frac{1}{2} (\sigma^3_l + 1).
\end{equation}
This preserves the anti-commutation relations. We use the (fermion) occupation number basis ($n_x \ket{0} = 0,
n_x \ket{1} = \ket{1}$, with $0$ denoting an unoccupied and $1$ an occupied site) to construct ${\cal Z}_{\rm f}$ by
splitting the Hamiltonian into different parts and invoking the Suzuki-Trotter formula. With some algebra to keep track
of the negative signs, the transfer matrix between the adjacent time-slices can be expressed through the following
$4 \times 4$ plaquette interactions (the state labels at sites $x$ and $y$ denote $\ket{00}, \ket{01}, \ket{10}, \ket{11}$):
\begin{equation}
\begin{aligned}
{\rm e}^{-S[n_1,n_2,n_3,n_4]} &= \bra{n_1 n_2} {\rm e}^{\epsilon \frac{{\rm t}}{2}(c^\dagger_x c_y + c^\dagger_y c_x) - \epsilon V (n_x - \frac{1}{2}) (n_y - \frac{1}{2}) } \ket{n_3 n_4} \\
                              &= {\rm e}^{\frac{\epsilon V}{4}}
  \begin{bmatrix}
 {\rm e}^{-\epsilon V/2} & 0 & 0 & 0 \\
 0 & {\rm cosh}(\epsilon {\rm t}/2) & \Sigma {\rm sinh} (\epsilon {\rm t}/2) & 0 \\
 0 & \Sigma {\rm sinh} (\epsilon {\rm t}/2) & {\rm cosh}(\epsilon {\rm t}/2) & 0 \\
 0 & 0 & 0 & {\rm e}^{-\epsilon V/2} \\
  \end{bmatrix}
\end{aligned}
\end{equation}
The structure of the transfer matrix is very similar to the one of the quantum spins, the main difference is the string
operator $\Sigma = \sigma^3_{l+1} \cdot \sigma^3_{l+2} \cdots \sigma^3_{m-1}$ between the sites $x = l$ and $x + \hat{i} = m$,
where $l < m$ in the lattice ordering. This allows the rewriting of ${\cal Z}_{\rm f}$ in the occupation number basis
$[\{ n_x = 0,1 \}]$ as
\begin{equation}
{\cal Z}_{\rm f} = \sum_{\{ n \}} {\rm sign} [\{n\}] {\rm e}^{-S[\{ n\}]}.
\end{equation}
\begin{figure*}
\includegraphics[width=1.0\textwidth]{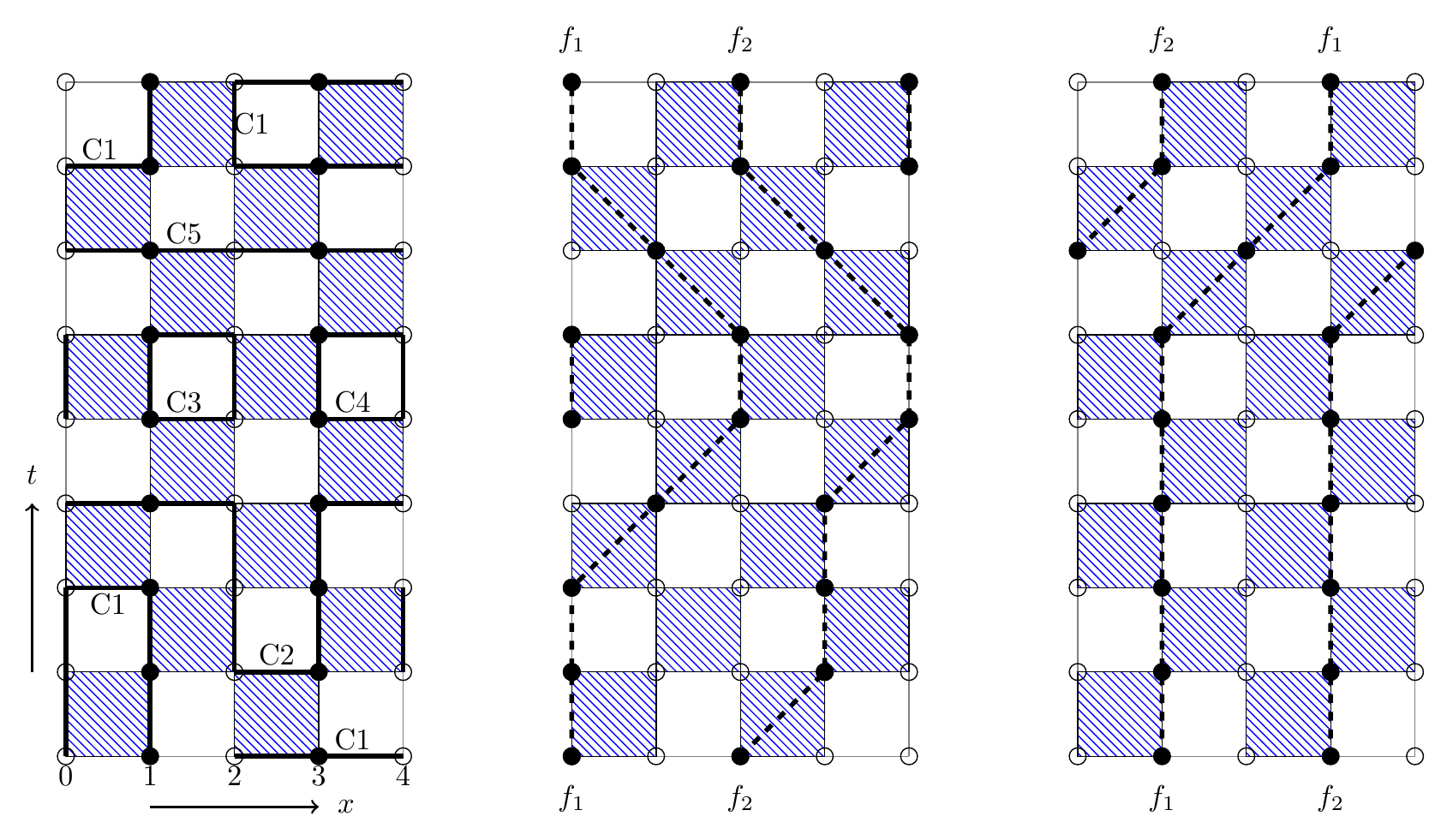}
\caption{Example configurations which contributes to ${\cal Z}_{\rm f}$. ({\bf left}) The two fermions $f_1$ and $f_2$
are static during their entire Euclidean time-evolution, and trace out vertical worldlines. For $t=J$, only $A$
(vertical) and $B$ (horizontal) bonds are required. The figure shows {\it one} bond configuration. The oriented
clusters are constructed by starting from a single point and following the bonds -- upwards (downwards) for a filled
(empty) site for $A-$breakups, sideways for a filled (empty) site touching $B-$breakups. This choice of breakups gives
rise to 5 different clusters, as marked in the figure. ({\bf middle}) On flipping clusters marked 1, 3 and 4 we obtain
a completely different world-line configuration, but one where the fermions do not exchange positions. ({\bf right})
Flipping cluster 5 gives rise to a configuration where the fermions interchange positions, and hence due to the
anticommutation relations, this configuration has an overall negative sign compared to the one on the left. The cluster
5 is a {\bf meron}.}
\label{fig2}
\end{figure*}
Some example configurations which contribute to ${\cal Z}_{\rm f}$ are shown in Fig. \ref{fig2}. Without the sign factor,
the weights of the resulting bosonic system  are identical to that of the XXZ-Hamiltonian:
\begin{equation}
H =  \sum_{x,\hat{i}=1,2} \left[ {\rm t} ( S^1_x \cdot S^1_{x+\hat{i}} + S^2_x \cdot S^2_{x+\hat{i}}) + {\rm V}  S^3_x \cdot S^3_{x+\hat{i}} \right],
\end{equation}
which reduces to the anti-ferromagnet for ${\rm t} = {\rm V} = J$. The ${\cal Z}_{\rm f}$ is decomposed in terms of
bonds, in addition to spins:
\begin{equation}
{\cal  Z}_{\rm f} = \sum_{[n,b]} {\rm sign} [n,b] ~ {\rm exp} (\left[ -S[n,b] \right])
\end{equation}
Interestingly, in the limit ${\rm t}={\rm V}$ the only breakups that are needed are the $A$ and the $B$ breakup, with same
probabilities as derived in Fig. \ref{fig1} (right). Fig \ref{fig2} shows a typical fermion configuration,
and a particular set of breakups, and how clusters are identified. Cluster flips are operations
$\{n_k \leftrightarrow 1-n_k | n_k \in {\cal C}_i \}$, where ${\cal C}_i$ is the i-th cluster. Cluster flips
give rise to new worldlines significantly different from their parent ones, but carry the same weights.

The {\it crucial consequence} of these cluster rules is that the sign of the whole configuration factorizes into
a product of the signs associated with each individual cluster. An example is already seen in Fig \ref{fig2} (right).
The clusters which can change the sign of the configuration are called {\bf merons}. It is possible to reach a
{\bf reference configuration} by flipping clusters appropriately. In this example, the configuration in Fig \ref{fig2} (left)
is the reference configuration, which has a positive sign (${\rm sign}_{{\cal C}_i}=1$) for all clusters i. When a
cluster is flipped, its contribution to sign is ${\rm sign}_{{\cal C}_i} = 1 (-1)$ if the quantity $N_w + N_h/2$ is
odd (even). $N_w$ is the temporal winding number of the cluster, while $N_h$ is the number of spatial hops \cite{Chan1999}.
The meron concept allows exact pairing of even and odd signs:
\begin{equation}
\begin{aligned}
{\cal Z}_{\rm f} &= \sum_{[n,b]} \overline{{\rm sign} [b]} ~ {\rm exp} (-S[n,b]) =  \sum_{[b], {\rm zero-meron}} 2^{N_{\cal C}} ~ {\rm exp} (-S[b]);\\
\overline{{\rm sign} [b]} &= \frac{1}{2^{N_{\cal C}}} \sum_{\rm cluster flips} {\rm sign} [n,b].
\end{aligned}
\end{equation}
$\overline{{\rm sign} [b]} = 0$ if at least one of the clusters is a meron. Further, the existence of the
reference configuration with a positive Boltzmann weight, guarantees that the unpaired configurations all
carry positive weight, and can be reached by flipping clusters from an initial configuration which has zero-merons.
The Monte Carlo sampling proceeds by generating configurations which only lie in the zero-meron sector. If a
proposed flip gives rise to meron cluster, it is rejected. Thus, one is able to completely solve the sign problem
with the meron concept.

Let us emphasize that for generic interactions the meron concept does not hold, and whether a cluster is a meron
depends on the orientation of other clusters. Therefore, much more (exponential) computational effort is needed
to identify the merons. However, there is a large class of interactions which can be solved with the meron concept,
and we refer the reader to \cite{Chan2002} for a review.

\section{Pure Gauge Theories}
Cluster algorithms for gauge theories run into problems since the relevant variables that need to be updated get
frustrated, and very often the clusters occupy the entire volume. The resulting algorithm is not much better than
a local update algorithm. One direction of research tried to identify the relevant degrees of freedom, which may
or may not be identical to the degrees of freedom in which the action is constructed. For example, for a $\phi^4$
theory, the $\phi$ field can be separated into an Ising ($Z_2$) variable and a modulus variable. The $Z_2$ degrees
of freedom were updated using a cluster algorithm while the modulus of the $\phi$ field was updated with a local
update algorithm \cite{Brow1989}, which resulted in an efficient algorithm with a low ($z < 1$) dynamical exponent.
Similar methods have been applied to the $SU(2)$ lattice gauge theory \cite{Ever1991,Kerl1994} targeting the Polyakov
loop as the embedded degree of freedom. While some success on the theories with one and two time-slices was reported,
these methods do not work well for larger lattices.

\begin{figure*}
\includegraphics[width=0.5\textwidth]{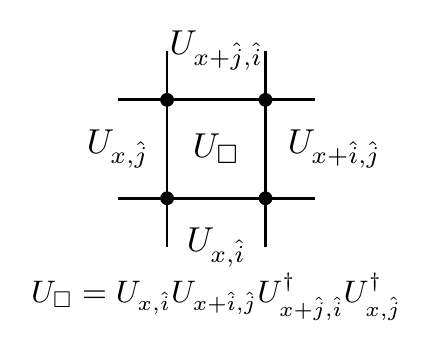}
\vspace{0.5cm}
\includegraphics[width=0.4\textwidth]{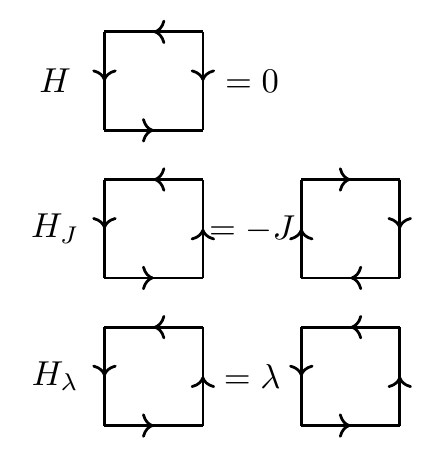}
\caption{({\bf left}) Layout of the plaquette, and the four spin interaction. ({\bf right}) The action of the
operators in the Hamiltonian on the possible plaquette states. The $U_\Box (U^\dagger_\Box)$ term flips a clockwise
(anti-clockwise) oriented plaquette to an anti-clockwise (clockwise) oriented one, and annhiliates the remaining
14 possible states. This is the $J-$ term of the Hamiltonian, and is an off-diagonal operator. The $\lambda-$term
counts the total number of flippable (both in the clockwise and anti-clockwise orientation) plaquettes.}
\label{fig3}
\end{figure*}

A completely different formulation of gauge theories, the quantum link model formulation \cite{Horn1981,Orla1990,Chan1997},
which realizes continuous gauge symmetries with finite dimensional Hilbert spaces and generalizes the Wilson
construction of lattice gauge theories, has recently been more amenable to simulation with cluster algorithms.
As an illustrative example, we will describe the $U(1)$ quantum link model (QLM). The degrees of freedom of this
Abelian lattice gauge theory are quantum links, $U_{x,\hat{i}}$ and electric fluxes, $E_{x,\hat{i}}$, both of which
are operators defined on the bonds connecting neighboring sites, $x$ and $x+\hat{i}$ and $\hat{i}=1,2$ in $d=2$.
These operators satisfy canonical commutation relations:
\begin{equation}
[E_{x,\hat{i}},U_{y,\hat{j}}] = U_{x,\hat{i}} \delta_{xy} \delta_{ij};~ [E_{x,\hat{i}},U^\dagger_{y,\hat{j}}] = -U^\dagger_{x,\hat{i}} \delta_{xy} \delta_{ij};~[U_{x,\hat{i}},U^\dagger_{y,\hat{j}}] = 2E_{x,\hat{i}} \delta_{xy} \delta_{ij},
\end{equation}
The operators $E, U$ can be chosen to be the components of a spin-$S$ object: $U = S^+, U^\dagger = S^-, E = S^3$.
The parameter $S$ can be thought of regulating the local Hilbert space dimension. In the limit $S \to \infty$
\cite{Schl2000}, these models reproduce the Hamiltonian formulation of Wilson gauge theories \cite{Kogu1974}.
The Hamiltonian we will be interested in only contains operators which commute with the Gauss' Law:
\begin{equation}
 H = -J \sum_\Box \left( U_\Box + U^\dagger_\Box \right)+ \lambda \sum_\Box \left( U_\Box + U^\dagger_\Box \right)^2 ; G_x = \sum_i \left(E_{x,x+i} - E_{x-i,x} \right)
\end{equation}
In addition, we could have included any function of the electric fluxes, and in particular, the kinetic energy of
the gauge fields $\sum_{x,i} E^2_{x,i}$. When the quantum link operators are considered in the spin-$\frac{1}{2}$
representation, this model has several interesting physical applications. In $(2+1)-$d, the physics of this model
is relevant for the understanding the behavior of low-temperature frustrated magnets \cite{Shan2004}, and spin-liquid
phases \cite{Herm2004,Bale2010}. A close cousin of this model is the {\it quantum dimer model} (QDM), especially
well-known in condensed matter physics as a toy model to describe certain aspects of superconductivity \cite{Rokh1988,Sach1989,Moes2007}.
Therefore, we continue with the gauge links in the spin-$\frac{1}{2}$ representation, and hence ignore the kinetic
energy term of the gauge links, which is a constant. The local link Hilbert space is 2-dimensional, and in the
electric flux basis they are denoted by upward (downward) arrows for $E = \frac{1}{2} (-\frac{1}{2})$ for vertical
links, and right (left) arrow for $E = \frac{1}{2} (-\frac{1}{2})$ for horizontal links. The action of the Hamiltonian
on the electric flux states is shown in Fig \ref{fig3}.

\begin{figure*}
\includegraphics[width=0.4\textwidth]{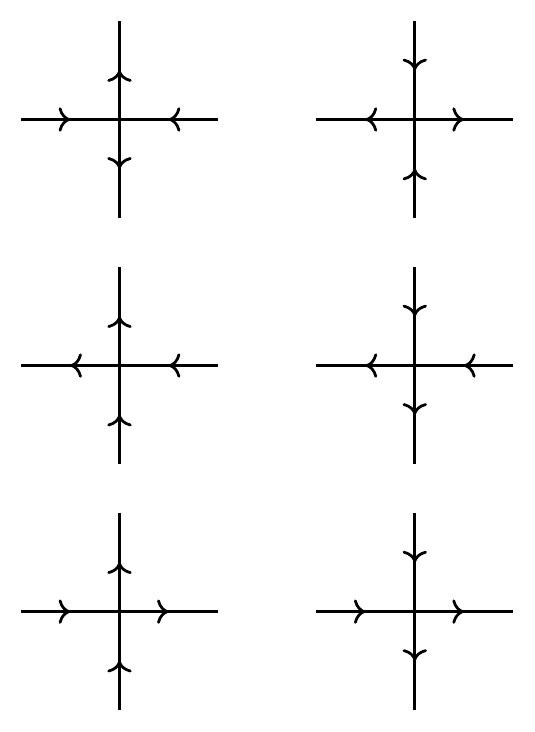}
\hspace{2cm}
\includegraphics[width=0.3\textwidth]{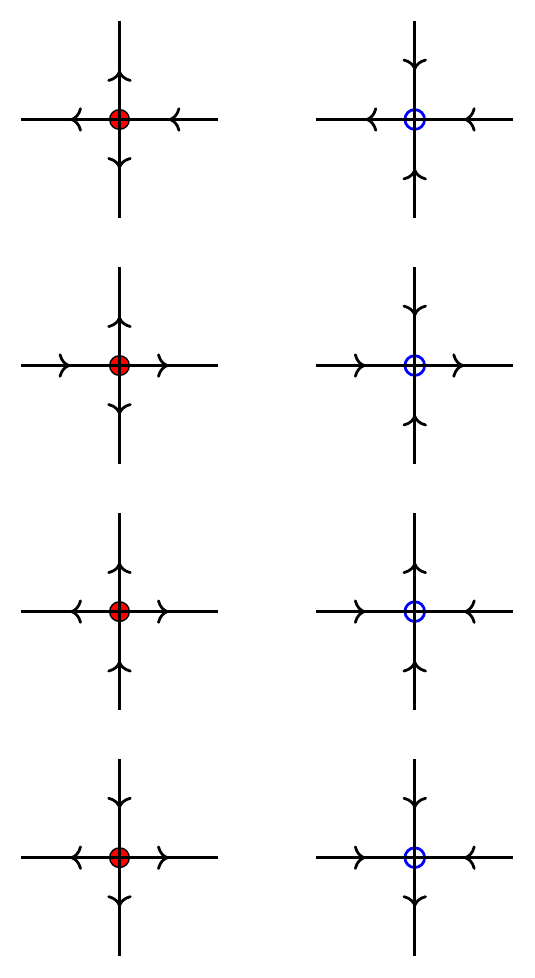}
\caption{({\bf left}) The Gauss' Law condition for the QLM, in $d=2$, this allows 6 possible states on each vertex,
while ({\bf right}) the Gauss' Law condition for the QDM in $d=2$ allows for 4 states for each vertex. The red circle
represents $Q = +1$, while the blue circle the $Q=-1$, which are distributed in a staggered fashion on the square lattice.}
\label{fig4}
\end{figure*}

Specifying the Gauss' Law further specifies which basis states can contribute to the partition function, and which not.
For the QLM, the Gauss' Law is chosen as $G_x \ket{\psi} = 0$, for every site $x$, for every eigenstate $\ket{\psi}$
of the Hamiltonian. The QDM chooses a different set of states, which are specified by
$G_x \ket{\psi} = (-1)^{x_1 + x_2} \ket{\psi}$. Physically, this implies that the vacuum selected by the QLM is
charge neutral at each vertex, while the QDM chooses a vacuum with staggered positive and negative unit charges. The
link states allowed at the vertex for the QLM is shown in Fig \ref{fig4} (left) and for the QDM in Fig \ref{fig4}
(right). The Gauss' Law generates the gauge transformations, which can be generically denoted as
$V = \prod_x {\rm exp}( i \theta_x G_x)$, where $\theta_x \in (0,2\pi]$ is the parameter of the transformation. The
Hamiltonian is invariant under this transformation: $\widetilde{H} = V^\dagger \cdot H \cdot V = H$, since
$[G_x, H] = 0$, for all x. Physically, $G_x$ labels different super-selection sectors of the theory which do
not mix under unitary dynamics.

\begin{figure*}
\includegraphics[width=0.5\textwidth]{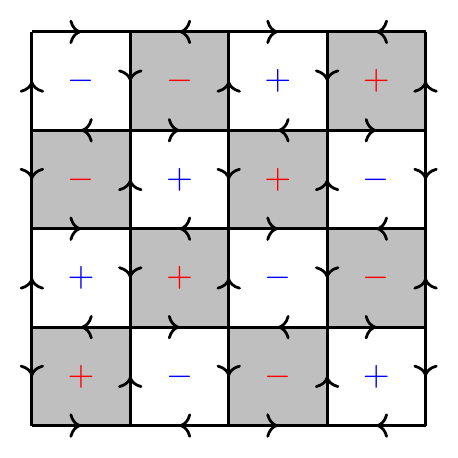}
\caption{ Mapping of an electric flux configuration (shown with arrows on the links) to a height configuration (shown
with $+$ and $-$ variables at the centers of the plaquettes). The $+$ and $-$ are placeholders for $0,1 (\pm \frac{1}{2})$
on A (B) sub-lattices. Every time a flux pointing right or upwards (corresponding to $E = \frac{1}{2}$) is crossed,
the height variable is changed, while it remains unchanged if a left or downward pointing flux (corresponding to
$E = -\frac{1}{2}$) is crossed. The configuration shown has all plaquettes flippable, and the corresponding height
variables are in their {\bf reference configuration}.}
\label{fig5}
\end{figure*}
The partition function of this model is defined by projecting the eigenstates of the Hamiltonian into a specified
Gauss' Law sector, ${\cal Z} = \mathrm{Tr} [\mathrm{exp} (-\beta H) {\cal P}_G]$. The projection operator
${\cal P}_G = \prod_x G_x $, therefore constrains the Hilbert space further. To construct the cluster algorithm, we
note that we can divide the Hamiltonian into two parts $H = H_A + H_B$, such that the {\it all} terms in each part
mutually commute. Thus, the Trotterization divides the lattice into even and odd sub-lattices, such that at a given
time-step only a single sub-lattice needs to be updated. For a fixed $\beta = \epsilon N$, we have a two-step transfer
matrix
\begin{equation}
{\cal Z} = {\rm Tr}~\left[(T_A T_B)^N P_{G} \right].
\end{equation}
The single plaquette transfer matrix can be expressed as
\begin{equation}
T_{\Box} = 1 + (U_\Box + U^\dagger_\Box) {\rm e}^{-\epsilon J \lambda} {\rm sinh}(\epsilon J) + (U_\Box + U^\dagger_\Box)^2 \left[ {\rm e}^{-\epsilon J \lambda} {\rm cosh} (\epsilon J) - 1 \right]
\end{equation}
The resulting transfer matrix is 16-dimensional (i.e. a $16 \times 16$ matrix), and it is easy to read off the
Boltzmann weights from this equation. For plaquettes which are not flippable, only the diagonal element is unity, and
all other off-diagonal elements are zero. For the two flippable plaquette configurations, the diagonal contribution
has the weight ${\rm e}^{-\epsilon \lambda} {\rm cosh}(\epsilon J)$ while the off-diagonal elements (which indicate
the plaquette flips) carry the weight ${\rm e}^{-\epsilon \lambda} {\rm sinh}(\epsilon J)$.

The idea behind the construction of the cluster algorithm is to {\it dualize} the gauge theory in $(2+1)-$d, which
gives rise to a $\mathbb{Z}(2)$ quantum height model in $(2+1)-$d. The dualization is an exact rewriting of the
partition function in terms of new degrees of freedom which are $\mathbb{Z}(2)$ degrees of freedom located at dual
sites. As shown in Fig \ref{fig5}, every flux configuration can be mapped to a height configuration. A configuration
of quantum height variables is assigned the values $h^A_{\tilde{x}} = 0,1$ and $h^B_{\tilde{x}} = \pm \frac{1}{2}$.
located at the dual sites $\tilde{x} = (x_1+ \frac{1}{2}, x_2+\frac{1}{2})$, and is associated with a flux
configuration
\begin{equation}
 E_{x,\hat{i}} = \left[ h^X_{\tilde{x}} - h^{X^\prime}_{\tilde{x}+\hat{i}-\hat{1}-\hat{2}}\right] {\rm mod}2 = \pm \frac{1}{2};~~X,X^\prime \in \{A,B\}.
\end{equation}
The modulo function acts by adding or subtracting 2 until the result is in the desired range $(-1,1]$. We note that
while for each height configuration there is exactly one flux configuration, the reverse is not true. It can be shown
that there are exactly two height configurations for each flux configuration. In addition we note that the mapping to
the height variables only works when (a) there are no charges at the lattice sites and (b) when charges $Q = \pm 2$
are located on the lattice sites. This explicitly indicates the $\mathbb{Z}(2)$ nature of the height variables.

\begin{figure*}
\includegraphics[width=1.0\textwidth]{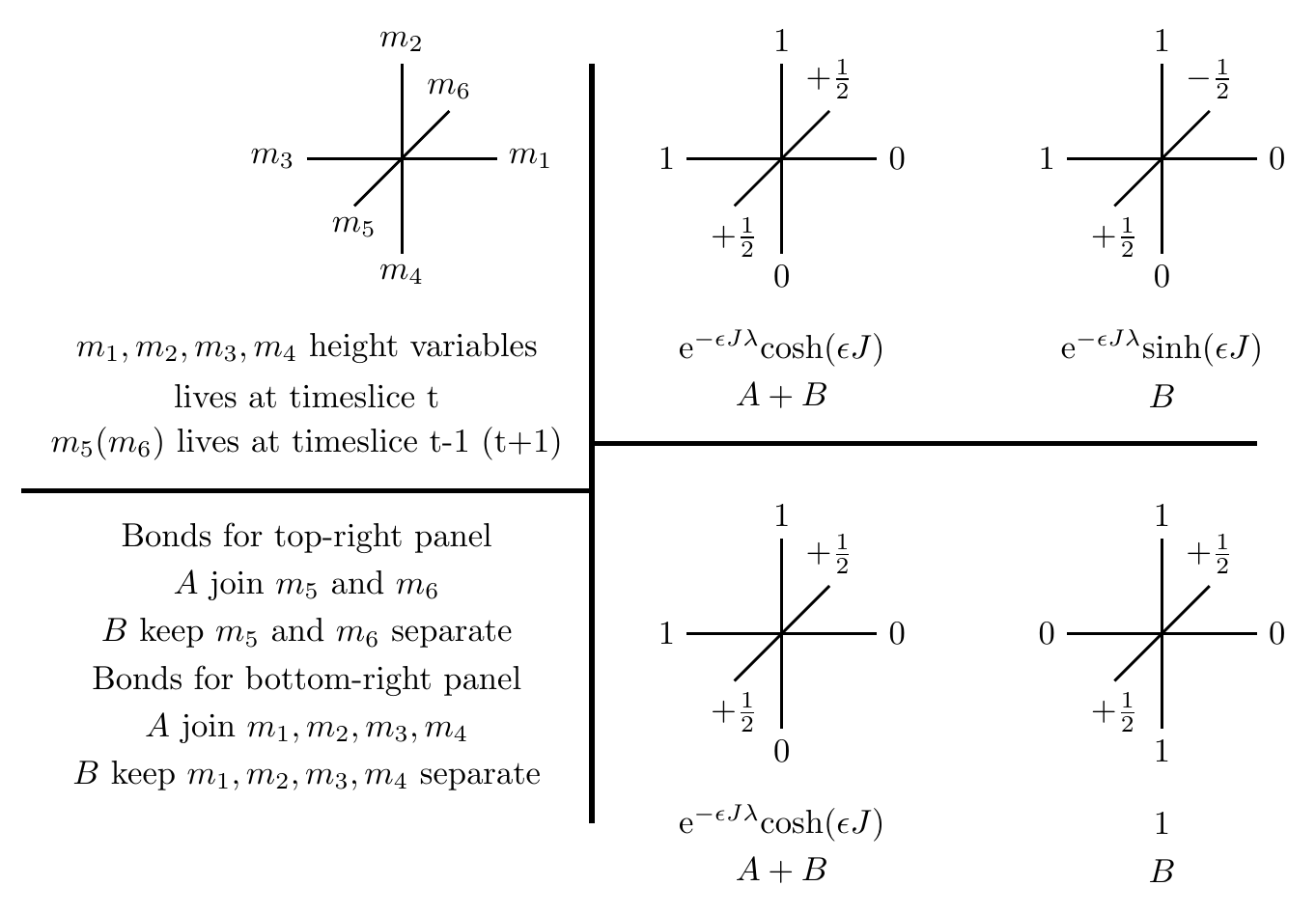}
\caption{({\bf left top}) Layout of the height variables across the timeslices t-1, t and t+1. ({\bf left bottom}) Two
kinds of bonds are enough to solve the linear equations: the A-breakup joins heights, while the B-breakup keeps the
heights separate. Different scenarios need to be considered --- out-of-plane (temporal) bonds, and in-plane (spatial)
bonds. ({\bf right top}) Cluster rules for temporal bonds: with $m_1,m_2,m_3,m_4$ in a reference configuration and
$m_5=m_6$, both A- and B-breakups can contribute, while if $m_5 \neq m_6$ then only the B-breakup contributes. To
satisfy detailed balance, we must have $P_A = A/(A+B) = {\rm e}^{-\epsilon J \lambda}/{\rm cosh}(\epsilon J)$ and
$P_B = {\rm tanh}(\epsilon J)$. In the case when $m_5 = m_6$ and $m_1,m_2,m_3,m_4$ are not in a reference configuration,
then we have to bind $m_5$ and $m_6$ with probability 1, otherwise a forbidden configuration will be generated.
({\bf right bottom}) Spatial breakups: if $m_5 \neq m_6$ then connect $m_1,m_2,m_3,m_4$ with probability 1, to prevent
the generation of a forbidden configuration. If $m_5=m_6$ then two different cases can occur: if
$m_1,m_2,m_3,m_4$ are in a reference configuration, then the Boltzmann weight is
${\rm e}^{-\epsilon J \lambda} {\rm cosh}(\epsilon J)$ and we can apply either the A, or the B-breakup. Otherwise,
the weight is 1 and only B breakup is applied ($m_5$ and $m_6$ are not bound together). Solving the
equations, we get $P_A = 1 - {\rm e}^{-\epsilon J \lambda}/{\rm cosh}(\epsilon J)$ and $P_B = 1 - P_A$.}
\label{fig6}
\end{figure*}
Before deriving the cluster rules, it is useful to understand the Boltzmann weights of the configurations in terms
of the height variables, as shown Fig \ref{fig6}. In terms of the height variables, the transfer matrix gives rise
to a 6-height interaction between the heights $m_1, \cdots, m_6$. Of these, $m_1,m_2,m_3,m_4$ live on a timeslice t,
$m_5$ lives on the timeslice t-1 and $m_6$ on t+1. Depending on the values of $m_1, m_2, m_3, m_4$, the height $m_5$
can undergo a transition to $m_6$ (corresponding to a plaquette flip), or not. In this language if
$(m_1,m_2,m_3,m_4) = (1,1,0,0)$ or $(0,0,1,1)$ the plaquette is flippable, and form the reference configurations.
The weights of these configurations and the derivation of the cluster rules is illustrated in Fig \ref{fig6} and
the figure caption.

With these cluster rules implemented in terms of the height variables, the super-selection sectors which have charges
$Q_x = \pm 1$ are never generated. However, it is still possible generate charges $Q_x = \pm 2$, since they are
compatible with the height representation. Therefore, in addition to the cluster rules for the Hamiltonian, the ones
for the Gauss' Law need to additionally implemented. To implement the Gauss' Law note that we need to consider four
height variables meeting around a site: the top-left and the bottom-right belong to a one sub-lattice (say A), and
top-right and bottom left belong to the other sub-lattice (B). It can be easily verified that if the site contains a
charge $\pm 2$, then both the height variables in {\it both} sub-lattices are locally out of their reference
configuration. Thus, this situation should always be avoided. This can be implemented as follows: when updating the
A sub-lattice at a given site, it is checked if the corresponding sites of the B sub-lattice are out of the reference
configuration. If this is the case then we bind the two height variables together, so that they are always flipped
together, and a forbidden configuration is not generated. If the relevant height variables in the B sub-lattice are
in a reference configuration then no additional bonds are put on the A sub-lattice heights. An identical procedure is
followed while updating the other sub-lattice. It is sufficient to check this on a single timeslice, since the Gauss'
Law commutes with the Hamiltonian.

We emphasize that this cluster algorithm for the gauge theory is somewhat different from the usual type of cluster
algorithms since clusters are separately built on each sub-lattice depending on the values of frozen height variables
in the other sub-lattice. In the next step, the frozen sub-lattice is updated. This algorithm was first implemented
in \cite{Bane2013} to study the physics of the U(1) QLM and it uncovered new phases of confined gauge theories which
break translation and charge conjugation symmetry \cite{Bane2013A}.

It is interesting to note that, even though this algorithm is very efficient on the QLM, extending this to the QDM
is not as useful. Since the Hamiltonian is the same, the same cluster rules can be applied. The problem is however
with the different Gauss' Law, which now frustrates the reference configuration. Due to the absence of the reference
configuration in the chosen basis of the quantum height model, the clusters grow quite large to occupy about $~80-90 \%$
of the space-time volume, which makes the cluster algorithm not much better than a local update Metropolis algorithm.
However, a combination of both these algorithms was used to study the physics of the QDM \cite{Bane2014,Bane2015}
and reliably extract the phase diagram at zero temperature. For the QDM certain other computational methods have
been reported \cite{Oake2018,Yan2018}, including certain novel effective theory approaches \cite{Herz2019}.

\section{Conclusions and Outlook}
Cluster algorithms are extremely efficient tools which are very useful to speed up the Monte Carlo sampling of strongly
interacting systems wherever they are applicable. These algorithms act by placing bonds between degrees of freedom
which are correlated, and build up patches in configuration space which can be {\it flipped}. The flipping process
gives rise to a new configuration which is very different from the parent one, but has the same Boltzmann weight.
Thus, the correlation in between the subsequent Monte Carlo configurations are removed, since the clusters themselves
are physical degrees of freedom. In the case of fermions, configurations often come with negative signs associated
with fermions interchanging their respective positions. A novel cluster algorithm --- the meron algorithm, is able
to analytically cancel clusters which are responsible for the negative worldlines, and only operate in the Hilbert
space which contribute non-trivially to the partition function. Cluster algorithms for gauge theories are even more
challenging than other systems since one has to identify proper degrees of freedom which can be used to build clusters
that do not occupy the whole space-time volume. In the case for certain quantum link models, this has been achieved
by dualizing the original Hamiltonian, and rewriting it exactly in terms of dual quantum height variables. The resulting
cluster algorithm is very efficient. Further, for gauge theories, the Gauss' Law needs to be implemented, which
may or may not be possible with the cluster rules that allow the sampling the full Hilbert space. For example, in
the case of QLM where the vacuum is charge neutral, this is not a problem, but the QDM which has a staggered background
charge this does not work so well, and the clusters become large. This can be understood as the lack of a reference
configuration for the QDM, which is an essential ingredient for the success of a cluster algorithm.

Demanding the existence of reference configurations, it is possible to construct models which are guaranteed to have
efficient cluster algorithm. Such Hamiltonians are often called {\it designer Hamiltonians} \cite{Kaul2013}, and can be studied in
any space-time dimensions. Since the exact form of the microscopic interactions is unimportant to study physics
associated with breaking of symmetries across thermal or quantum phase transitions, thanks to universality, cluster
algorithms can be used to study a wide range of physical phenomena in naturally occurring strongly correlated systems.

In the past few years, there has been renewed interest with cluster and meron algorithms. In the context of classical
spin models, the meron algorithm is being used to study the topological charge and the vorticity properties of non-linear
sigma models \cite{Biet2019}. For lattice fermions, the meron algorithm has been used to study a Hamiltonian spin-half
lattice fermions that displays symmetry enhancement for certain coupling regimes \cite{Liu2020}. In the context of
gauge theories, the dualization technique has been extended to a non-Abelian $SU(2)$ quantum link model on the honeycomb
lattice and an efficient cluster algorithm has been constructed in terms of the dual height variables \cite{Bane2017}.
This cluster algorithm operates on a four sub-lattice structure, where the clusters are built using the heights on one
sub-lattice, while the heights on the other three sub-lattices decide the nature of bonds. This study identified more
crystalline confined phases in non-Abelian theories, as well as showed fractionalization of a $Z(2)$ flux.

Another very novel and interesting development with cluster algorithms is the simulation of strongly correlated
systems in {\bf real-time and far out of equilibrium}. The first such study considered interesting initial states
which are in contact with a thermal reservoir, and are completely driven via dissipation. The authors of \cite{Bane2014A}
devised a cluster algorithm to simulate the Lindblad evolution in an anti-ferromagnetic initial state. More extensive
studies revealed what kinds of measurements give rise to sign problems, and which measurements did not \cite{Hebe2015}.
Transport phenomena in strongly correlated spin-$\frac{1}{2}$ driven via Lindblad dynamics was studied using
cluster algorithms in \cite{Bane2015A}.

\begin{acknowledgements}
This article is written in the memory of the late Pushan Majumdar. Pushan taught me the intricacies of the multi-level
algorithm \cite{Lues2001,Maju2003}, a highly efficient Monte Carlo method which greatly accelerates the simulation of
pure Wilson-type gauge theories. Besides, I would like to acknowledge Pushan for the many discussions on physics, his
advice on parallel and GPU programming that we have had, and for the many cups of coffee he would prepare after lunch.

I would like to thank Shailesh Chandrasekharan and Uwe-Jens Wiese for teaching me so much about cluster and meron
algorithms. I would also like to acknowledge Saumen Datta, Fu-Jiun Jiang, Kieran Holland, Emilie Huffman, Ferenc Niedermayer,
Stefan Schaefer, Arnab Sen, Rainer Sommer, Urs Wenger, and Ulli Wolff for various discussion on improved algorithms.
\end{acknowledgements}

\end{document}